# Dielectric Investigation, Piezoelectric Measurement and Structural Studies of Strontium-doped, modified PMS-PZT Piezoelectric Ceramics

Kumar Brajesh[1,4], Rajyavardhan Ray[2,3a], A. K. Himanshu[3a,*], Yashwant Kumar[3b], Rajeev Ranjan[4], N. K. Singh[1], S. K. Bandyopadhayay[3a], T P Sinha[5]

[1]Department of Physics, Veer Kunwar Singh University, Ara, PIN- 802301, Bihar, India

[2]Department of Physics, Indian Institute of Technology Kanpur, India

[3]Nanostructured & Advanced Materials Laboratory[a] & PSI[b], Variable Energy Cyclotron Centre. DAE, 1/AF Bidhannagar, Kolkata, 700064, India.

[4]Materials Engineering, Indian Institute of Science, Bangalore -560012, India

[5]Department of Physics, Bose Institute, 93/1, Acharya Prafulla Chandra, Road, Kolkata 700009, India

**Abstract:** The structure and dielectric properties of 6% $Sr^{2+}$-substituted, modified Lead Zirconium Titanate (PZT) piezoelectric ceramic, with composition $[Pb_{0.94}Sr_{0.06}][(Mn_{1/3}Sb_{2/3})_{0.05}(Zr_{0.54}Ti_{0.46})_{0.95}]O_3$ and lead manganese antimonite as an additional dopant, synthesized by ceramic route have been investigated in a frequency range from 50Hz to 1MHz and in a temperature range between room temperature and 633K. The scanning electron micrograph of the sample taken at the room temperature confirms the formation of a fairly homogenous structure and well-formed grains with sharp boundaries. From the Rietveld analysis, it was found that the structure of the material is tetragonal with space group P4mm. The scaling behavior of imaginary part of the electric modulus suggests that the relaxation is described by the same mechanism at various temperatures. The values of the electromechanical coupling factor ($k_p$), the high mechanical quality factor ($Q_m$) and the piezoelectric coefficient ($d_{33}$) of the synthesized sintered & poled ceramics gives a good set of values for the piezoelectric properties when compared with related materials, useful for various interesting piezoelectric device applications.

*Corresponding authors. Tel.: +91 33 23371230 (Ex:2408), Mobile.: +91 9883239177 Fax.: +91 33 23346871, Email : himanshu_ak@yahoo.co.in, akh@vecc.gov.in ( A. K. Himanshu).

# 1. Introduction

The groups of materials with $ABO_3$ −type perovskite structure are very important due to their rich physical and chemical properties, leading to attractive electrical and magnetic behavior for possible technological applications such as advanced sensor, actuator technology and piezoelectric transformers [1-6]. In particular, Lead Zirconate Titanate (PZT), especially for x=0.52 when PZT is close to the morthotropic phase boundary (MPB) with the formula $Pb(Zr_xTi_{1-x})O_3$, have received special attention, [7]. The MPB is an almost temperature-independent phase boundary that separates two ferroelectric phases *viz.* the tetragonal and rhombohedral crystal structure. Close to MPB, PZT displays exceptional piezoelectric and ferroelectric properties, the properties being high values of the electromechanical coupling coefficient $k_p$, the mechanical quality factor $Q_m$, and the piezoelectric strain coefficient $d_{33}$. PZT is rarely used in it's pure form. Numerous attempts, e.g. doping the A- or B-site, have been undertaken to tailor its properties for specific applications, leading to modified PZT structures. Depending on the nature of the dopant, these modified PZT composites are broadly classified as hard (donor dopant) and soft (acceptor dopant) PZT [8-11] briefly discussed below.

The hard and soft PZTs have their own advantages. While the hard PZT has relatively low dielectric loss factor and high mechanical quality factor ($Q_m$), the soft PZT has a relatively high piezoelectric constant along with a high coupling-coefficient ($k_p$). Usually hard piezoelectric ceramics are found to be useful for transformer applications because of their high mechanical quality factor ($Q_m$). Modified PZT ceramics have also found applications in high power and transmitting components which demand high mechanical, dielectric and piezoelectric properties. In order to obtain ceramics which combine the advantages of both hard and soft PZTs, different modifications have been investigated [9-11]

Among hard additives, $Mn^{2+}$ and $Fe^{3+}$ generate $O^{2-}$ vacancies. While $Mn^{2+}$ (or $Fe^{3+}$) cation replaces $Zr^{4+}$or $Ti^{4+}$ ions on the B-site, $Sr^{2+}$ cation replaces the $Pb^{2+}$ ions on the A-site. The O-site vacancies, on the other hand, lead to contraction of the grain body. At the same time, defect-complexes, which consist of impurity ions and O-site vacancies along with the domain boundaries, are generated to induce long–range polar order. Owing to these reasons, motion of the domain wall gets harder. Consequently, the weak-field dielectric and piezoelectric constants, and the weak-field dielectric losses reduce whereas the coercive field and mechanical quality factor increases. Modified PZT structures, esp. PMS-PZT are being extensively studied and have shown rexalor behavior.

Sr-doped PZT materials (known as PSZT) have recently been studied as a function of Sr content [12-15]. It is found that PSZT undergoes a diffuse type ferroelectric phase transition which depends on the substitution of Sr for Pb and that the diffusivity increases with increase in Sr doping. This allows the PZT materials to be used in a wider variety of device application [12]. The Presence of the Sr leads to straining of the PZT lattice due to its smaller size [13]. Theoretical and experimental studies by Nasar et al.[14] have also found that Sr doping promotes a structural change from tetragonal ferroelectric PZT to tetragonal and rhombohedral ferroelectric phases (MPB), Sr-PZT. Decrease in tetragonal phase is due to the appearance of rhombohedral phase of Sr-PZT. Formation of rhombohedral phase is caused by an increase of compositional fluctuation with a consequent coexistence of both these phases. This is accompanied by strong decrease in (c/a) ratio in the tetragonal phase due to minor repulsion between $Sr^{2+}$ and $O^{2-}$ orbitals in the structure. This leads to alteration of dipole formation due to minor $Zr^{4+}/Ti^{4+}$ ion dislocation, eventually leading to increase in the value of the planar coupling coefficient $k_p$. However, no relaxor behavior has been found.

In this paper, we have, therefore, concentrated on a hard piezoelectric ceramic composition $[Pb_{0.94}Sr_{0.06}][(Mn_{1/3}Sb_{2/3})_{0.05}(Zr_{0.54}Ti0_{.46})_{0.95}]O_3$ (Sr6-PMS-PZT) as a case study to explore its various dielectric and piezoelectric properties for piezoelectric transformer application. It must be noted that a very similar composition has been considered earlier [16] where 2% Sr doping was considered in the presence of $(Mn_{1/3} Sb_{2/3})_x$ at the B-site (x varying between 0.025 and 0.1). The MPB boundary was found near x=0.05. The synthesis of PZT system is, in general, not without the presence of an unwanted pyrochlore phase and Sr doping is known to decrease the pyrocholre content in PZT and related systems [17]. Therefore, the motivation for 6% Sr doping is two-fold: (i) in obtaining pyrochlore-phase free PMS-PZT system, and (ii) It is also known that the piezoelectric properties of $Sr^{2+}$-substituted PZT are most pronounced for x = 0.06, which is also the basis of the commercial PZT4 [13,18,19]. Furthermore, to the best of our knowledge, no relaxation studies have been carried out for the Sr-doped PMS-PZT structures.

The present work, therefore, aims to study the impact of 6% Sr doping on the A-site, along with the presence of (Mn,Sb) (fixed concentration) doping on the B-site (PMS-PZT) on the dielectric and piezoelectric properties of modified PZT system. We have carried out a careful study of the crystal structure, and the dielectric and piezoelectric properties. It is found that the dominant phase in the composition is tetragonal with space group P4mm ($\chi^2 = 1.50$). The frequency dependent electrical data has been analyzed in the framework of conductivity and electric modulus formalisms. The Cole-Cole model has been used to study the dielectric relaxation [20, 21] of the synthesized material.

## 2. Experiment

Stoichiometric amounts of the $PbCO_3$ and $SrCO_3$ powders were taken for the preparation of $[Pb_{0.94}Sr_{.06}]CO_3$. Saturated solution of ammonium carbonate was added to the solution of $PbCO_3$ and $SrCO_3$ in dilute Nitric acid to obtain the precipitate of $[Pb_{0.94}Sr_{.06}]CO_3$. The precipitate was then washed with distilled water until ammonia was removed and later dried in an oven. In order to obtain the preparation of the final composition Sr6-PMS-PZT, $[Pb_{0.94}Sr_{.06}]CO_3$, and $MnSb_2O_6$, $ZrO_2$ (99% purity) and $TiO_2$ (99.5% purity) were taken in the desired stoiciometry. Details regarding the synthesis has been discussed in out previously reported work [22].

The XRD studies (using Rigaku miniflex) were performed for the sintered specimen to determine the structural details of the material. For the characterization of electrical behavior, the sintered ceramic pellets were electroded (Hioki LCR meter electrodes) using silver paste. After applying the paste the pellets were dried at about $150^0C$ in an oven. The silver paste coated pellets were fired at $500\ ^0C$ for five minutes. The microstructure of sintered specimen was then examined using scanning electron microscope (FEI Quanta 200). Sintered ceramic bodies of Sr6-PMS-PZT were converted into piezoelectric bodies by poling. In order to determine the values of the planar electromechanical coupling coefficient ($k_p$) and the Mechanical Quality Factor ($Q_m$) of poled ceramic samples, the resonance and anti-resonance frequency measurements were carried out using a sample holder, resistance box and shielding furnace.

Rietveld refinement was carried out using the XRD data with the help of the FULLPROF program [23]. The background was fitted with a 6-cofficient polynomial function, while the peak shapes were described by pseudo-Voigt profiles. In all the refinements, the scale factor, the lattice parameters, the positional coordinates (x,y,z) and the thermal parameters were varied. Occupancy parameters of all the ions were kept fixed during the refinement. No correlation between the positional and thermal parameters was observed during refinement and, as such, it was possible to refine all the parameters together.

## 3. Results and discussion:

Fig. 1 shows the X-ray diffraction (XRD) pattern of the present composition and all the reflection peaks of the X-ray profile are indexed. The structure is tetragonal (space group P4mm). The $Pb^{2+}/Sr^{2+}$ ions occupy the *1a* sites at (0,0,z), $Mn^{2+}/Sb^{5+}/Zr^{4+}/Ti^{4+}$ and $O_I^{2-}$ occupy *1b* sites at (1/2,1/2,z), and $O_{II}^{2-}$ occupy *2c* sites at (1/2,0,z). For the refinement, the initial values of the lattice parameters were obtained from our XRD data by the least squares method, whereas the values of the structural parameters were taken from Noheda *et al.* [24] In this structure, $Pb^{2+}/Sr^{2+}$ coordinates were fixed at (0,0,0) in our refinement. Fig. 2 shows the observed, the calculated and the difference profiles for the refined structure, after having removed the peaks corresponding to the Cu-k$\alpha_2$ wavelength ($\lambda$=1.5406Å). The fit is found to be remarkably good. The refined structural parameters and the corresponding positional coordinates of this composition are given in Table 1.

To analyze the structure of the present composition, we have concentrated on the 200, 220, 222 pseudo-cubic XRD profiles as shown as the inset of Fig. 1. The structure of pure PZT is reported to be either rhombohedral or tetragonal or coexistence of these two phases [14]. It has also been shown that a monoclinic phase with space group Cm may coexist with the tetragonal phase in the MPB region [25]. For the rhombohedral structure, the 200 reflection peak appears as a singlet, where as the 220 and the 222 peaks appear as doublet with weaker reflection on the lower 2$\theta$ side. For a tetragonal structure, while the 200-peak is a doublet with weaker reflection on the lower 2$\theta$ side, the 220 peak is also a doublet with weaker reflection on the higher 2$\theta$ side whereas the 222 peak is a singlet. For a monoclinic structure, all the three reflection peaks *viz.* the 200, 220, and 222 peaks show splitting. For the present composition, we analyze this set of peak profiles. Since the 200 peak is a doublet with weaker reflection on the lower 2$\theta$ side, the 220 peak is also a doublet with weaker reflection on the higher 2$\theta$ side and the 222 peak is a singlet (see the inset of Fig. 1). Hence, we find that the structure of this composition is tetragonal, the space group being P4mm as obtained from the Rietveld refinement. Fig. 3 shows the SEM of the sample, indicating that the sample has a homogeneous phase with fairly uniform grains of size ≈ 5.9 µm.

In order to study the dielectric properties of the composition, we have studied the dielectric permittivity as a function of frequency and temperature. The angular frequency ω (=2πυ) dependent plots of the real (ε') and imaginary (ε") parts of complex dielectric permittivity(ε*) of the given composition at several temperatures ranging from room temperature to 523K are shown in Fig.4(a) & (b). A gradual decrease in ε'(ω) and presence of a broad peak in ε"(ω) is observed, implying relaxation in the entire temperature range considered.

The dielectric relaxation of the material can be analyzed in term of dipolar and conductivity relaxation mechanisms. The dipolar relaxation can be modeled by the Debye or the Cole-Cole equation. In dipolar relaxation process, relaxation phenomena are associated with a frequency dependent orientational polarization. At low frequency values, the permanent dipoles align themselves along the field and contribute fully to the total polarization of the dielectric. At higher frequencies, however, the variation in the field is too rapid for the dipoles to align themselves. Thus, their contributions to the polarization and, hence, to the dielectric permittivity can become negligible. Therefore, the dielectric constant ε'(ω) decreases with increasing values of frequency, which is a typical characteristic behavior for ferroelectric materials [26].

The peak position of ε"(ω) (centered at the dispersion region of ε'(ω)) in Fig. 4(b) shifts to a higher frequency with increasing values of temperature, indicating a strong dispersion. Also, there is an increase in the peak value of ε"(ω) with the increase in temperature, indicating an increase in charge carriers in the given material by thermal activation and, thus, showing a semiconducting behavior of the sample. The increase in the value of ε"(ω) [Fig. 4(b)] in low frequency region above 423K is due to dc conductivity [21,22]. Furthermore, it is apparent that the width of the loss peaks in Fig. 4(b) cannot be accounted for in terms of a mono-dispersive relaxation process but possibly through a distribution of relaxation times. One of the most convenient ways of checking the poly-dispersive nature of the dielectric relaxation is through the Cole-Cole model where the complex dielectric constant is known to be described by the empirical relation [27]:

$$\varepsilon^*(\omega) = \varepsilon'(\omega) - i\varepsilon''(\omega) = \varepsilon_\infty(\omega) + (\varepsilon_s - \varepsilon_\infty(\omega))/(1+(i\omega\tau)^{1-\alpha}), \quad \ldots\ldots(1)$$

where $\varepsilon_s$ and $\varepsilon_\infty(\omega)$ are the low- and high–frequency values of $\varepsilon'(\omega)$ respectively, $\tau = (1/\omega)$ is the relaxation time, and $\alpha$ is a measure of the distribution of relaxation times. Upon fitting the experimental data in Fig. 4(b) with Eq. (1), it is found that the experimental data almost exactly resembles (not shown) the values of $\alpha$ between 0.84 and 0.88 corresponding to the temperature ranging from 303k to 433k. A good agreement between the directly measured values of $\varepsilon''(\omega)$ and those calculated using Eq.(1) suggests that the relaxation process differs from the mono-dispersive Debye process, which corresponds to $\alpha = 0$.

Fig.5 shows the respective variation of $\varepsilon'(\omega)$ and $\varepsilon''(\omega)$ with temperature at various frequencies. At temperature far above $T_m$ and at two frequency values 100Hz & 1KHz, a monotonic increase in the value of $\varepsilon''(\omega)$, caused by electrical conduction, is observed. The phase transition involved is diffuse and the dielectric constant is markedly dispersive below the temperature $T_m'$ at which it peaks. The temperature $T'_m$ and $T''_m$, corresponding to the peaks in real ($\varepsilon'(\omega)$) and imaginary ($\varepsilon''(\omega)$) parts of the dielectric constant respectively, are not coincident which is a typical relaxor behavior.[22] In fact, the temperatures $T'_m$ and $T''_m$ are frequency dependent with $T''_m < T'_m$ and they increase with increasing values of frequency. The temperature dependence of the characteristic relaxation time is shown in Fig. 6, which satisfies the Arrhenius law:

$$\omega_m \tau_0 \exp(E_e / k_B T) = 1. \quad \ldots(2)$$

From the numerical fitting analysis, the value of the activation energy is $E_e \approx 0.11 eV$.

The dispersion (frequency dependence) of the dielectric response may be attributed to the relaxation of individual $TiO_6$ octahedral out of the network of ferroelectrically-coupled $TiO_6$ octahedrals. It has earlier been reported that isovalent doping at the A-site prevents the coupling between the ferroelectric networks formed by $TiO_6$ octahedra [21]. With increase in $Sr^{2+}$ content, there is a progressive breakdown of relatively large $Pb^{2+}$ rich regions into smaller regions. This disrupts the coupling and leads to decrease in volume of micro-polar regions. As a general rule, an addition of a non-ferroelectric component is known to cause smearing and shifting of the $T_m$ toward lower temperatures and promote the appearance of relaxor behavior, as also evident from Fig. 5(a).

We have also plotted the imaginary part (M″) of complex electric modulus (M*) as a function of frequency (angular) at several temperatures in Fig. 7. The frequency region below peak maximum ($M_m$) determines the range in which charge carriers are mobile on long distances. At frequencies above peak maximum, the carriers are confined to potential wells and, therefore, mobile for short distances. The value of $M″_m$ increases with increasing temperature and shifts to higher frequencies. At any temperature, the most probable relaxation time corresponding to the peak position in M″ versus ln ω plot satisfies the Arrhenius law with activation energy $E_M \approx 0.12$ eV, which is nearly same with the activation energy $E_e$, as shown in Fig.6.

The scaling behavior of M″ is shown in Fig.8, where we have scaled each M″ at different temperature values by $M″_m$ and corresponding frequency by $\omega_m$ ($\omega_m$ corresponds to the frequency of the peak position of M″ in the M″ versus ln ω plots in Fig.7). The overlap of the curves at different temperatures clearly indicates that the relaxation process is described by the same mechanism at various temperatures in the temperature range considered. A similar collapse of the M″(ω,T) data onto a single curve was reported earlier [22]. Fig. 8 shows the variation of the real part of ac conductivity as a function of frequency σ(ω) for given composition at different measuring temperatures. The conductivity shows a dispersion that shifts towards higher frequencies with an increase in temperature. It is seen from Fig. 9 that σ decreases with decreasing frequency. Extrapolation of this part towards lower frequency gives the value of dc conductivity $\sigma_{dc}$. Fig.10 shows the variation of normalized electric modulus and loss tangent, respectively (M″ / $M″_m$) and (tanδ/tan$δ_m$), as a function of logarithmic frequency measured at 333K for given composition. For delocalized or long-range conduction, the peak position of two curves should overlap. However, for the present system the M″/$M″_m$ and tanδ/tan$δ_m$ peaks overlap partially, suggesting contributions from both long-range and localized relaxation processes.

Sintered ceramic bodies of present composition were converted into piezoelectric bodies by poling. The values of planar coupling factor ($k_p$), piezoelectric constant ($d_{33}$) and mechanical quality factor ($Q_m$) are found to be 0.57, 281 pC/N and 1192 respectively, which is very encouraging and indicates a good performance of the given ceramics. We will now compare these values with some of high power piezoelectric materials reported in the literature. In this regard, we consider the following ceramic compositions:

Ceramics-A : $[Pb_{0.94}Sr_{0.06}]\,[(Mn_{1/3}Sb_{2/3})_{0.05}(Zr_{0.53}Ti0_{46})_{0.47}]O_3$ [22],

Ceramics-B: $0.05Pb(Mg_{1/3}Ta_{2/3})O_3\text{-}0.05Pb(Mn_{1/3}Sb_{2/3})O_3\ 0.9Pb(Zr_{0.52}Ti_{0.48})O_3$ [28],

Ceramics-C: $Pb(Mn_{1/3}Sb_{2/3})O_3\text{-}PbZrO_3\text{-}PbTiO_3$ [29],

Ceramics-D : $(Pb_{0.94}Sr_{0.06})(Ti_{0.47}Zr_{0.53})O_3$ [7],

and, Ceramics-E: $Pb_{0.98}\,Sr_{0.02}(Mn_{1/3}Sb_{2/3})_{0.05}\,(Zr_{0.5}Ti_{0.5})_{0.95}$ [16]

Table 2 shows a comparison of the piezoelectric parameter for these samples, clearly showing that the present composition has comparable values of the piezoelectric parameters, thus, being a promising candidate for high power piezoelectric device applications as also reported earlier for related systems [30,31]. Moreover, the considered ceramic compositions were found to be close to MPB boundary which was responsible for the performance of these ceramics. Comparable values of the present composition suggest that the present composition may also be lying close to the MPB boundary [32].

Although Ceramic E is structurally similar (tetragonal) to the present composition, its values of the piezoelectric parameters are slightly large due to smaller Sr content, as explained in the following. Generally, Mn ions, which exist in $Mn^{2+}$ and $Mn^{3+}$ states, in the perovskite structures[18] and enter the oxygenic octahedral center of the Perovskite system by substituting the B-site ions. Therefore, $Zr^{4+}/Ti^{4+}$ substitution at the B-site by Mn ion leads to oxygen vacancies. Radius of the Sr ion, on the other hand, is smaller but comparable to the Pb ions at the A-site, the respective values being 1.44 Å and 1.49 Å. Smallness of the Sr ion, together with the O vacancies created by Mn ions, leads to distortion of the tetragonal lattice through decreasing (c/a) ratio, eventually leading to a decreased unit cell volume in this phase. (This also allows for the formation of rhombohedral phase.) At the same time, comparable values of the ionic radii of Sr and Pb ion could reduce oxygen vacancies by forming A site vacancy-oxygen vacancy defect dipoles. This combined with the breaking of coupling between TiO6 and Pb ions, as mentioned above, leads to degradation of the ceramic properties. For large Sr content, some of the Sr ions get accumulated on the grain boundaries which hinder the domain wall movement, leading to further hardening and, therefore, may result in the further degradation of ceramic properties.

## 4. Conclusions

In conclusion, a Pyrochlore-phase free sample of $[Pb_{0.94}Sr_{0.06}][(Mn_{1/3}Sb_{2/3})_{0.05}(Zr_{0.54}Ti_{0.46})_{0.95}]O_3$ was synthesized by chemical route. The Rietveld refinement of the XRD data of the sample shows the composition has a tetragonal phase with P4mm structure, the lattice parameters being: a = b = 4.046838 Å and c = 4.0890656Å. The SEM indicates that the sample has a fairly homogeneous phase with uniform grains of size ≈ 5.9 μm. The capacitance and conductance of material are measured in a frequency range from 50Hz to 1MHz at various temperatures between 303K and 633K. A relaxation is observed in the entire temperature range as a gradual decrease in $\varepsilon'(\omega)$ and as a broad peak in $\varepsilon''(\omega)$. The frequency-dependent maximum in the imaginary electric modulus at various temperatures is found to obey an Arrhenius law with activation energy of 0.11eV. Scaling behavior of the imaginary part of electric modulus suggests that the relaxation process is described by the same mechanism at various temperatures. The temperature dependent dielectric studies indicate that the present composition is also a relaxor. The $k_p$, $Q_m$ and $d_{33}$ values of present composition has been compared with other related materials, indicating a strong promise of the present sample for various piezoelectric device applications.

## Acknowledgement

One of the authors (KB) acknowledges the financial help of the University Grants Commission, New Delhi, India [ref. No. 33-29/2007(SR)]. Authors AKH, SKB & YK are thankful to Prof. D. K. Srivastava, Director of VECC for his keen interest and encouragement in the XI-year plan project related work for piezoelectric device applications (PIC No. 11-R&D-VEC-5.09.2000).

[32] In fact, owing to this result and Ref [22], we have now explored this class of ceramics for variable Zr content, ranging from 0.53 to 0.60. We indeed find MPB boundary for Zr content =0.55. These results will be published separately.

**Table**

**1.** Refined structural parameters of $[Pb_{0.94}Sr_{.06}][(Mn_{1/3}Sb_{2/3})_{0.05}$ $(Zr_{0.54}Ti_{0.46})_{0.95}]O_3$ using tetragonal phase (space group: P4mm) model.

---

| Ions | Positional coordinates | | | Thermal parameters $U(Å^2)$ |
|---|---|---|---|---|
| | x | y | z | |
---

| $Pb^{2+}$ / $Sb^{2+}$ | 0.00 | 0.00 | 0.00 | $U_{11} = U_{22} = 0.04086(3)$ |
| | | | | $U_{33} = 0.00837(3)$ |
| $Mn^{2+}$ / $Sb^{5+}$ / $Zr^{4+}$ / $Ti^{4+}$ | 0.50 | 0.50 | 0.52698(4) | $U_{11} = U_{22} = 0.00184(2)$ |
| | | | | $U_{33} = 0.05037(1)$ |
| $O_I^{2-}$ | 0.00 | 0.50 | 0.58516(4) | $U_{iso} = 0.08200(1)$ |
| $O_{II}^{2-}$ | 0.50 | 0.50 | 0.07692 | $U_{iso} = 0.05916(3)$ |

---

a = b = 4.046838(4)Å, c = 4.090656(4) Å, $R_P$ = 3.43, $R_{wp}$ = 4.44, $R_{exp}$ = 3.62 and $\chi^2$ = 1.50

**2**. Comparison of piezoelectric properties of the given material with those of earlier reported high power piezoelectric ceramics.

| | Present Ceramics | Ceramics-A | ceramics-B | ceramics-C | ceramics-D | Ceramic E |
|---|---|---|---|---|---|---|
| $k_p$ | 0.57 | 0.53 | 0.55 | 0.60 | 0.52 | 0.64 |
| $d_{33}$ | 281 pC/N | 271 pC/N | 227 pC/N | 374 pC/N | 223 pC/N | 410pC/N |
| $Q_m$ | 1192 | 1115 | 1830 | 1250 | 500 | 1220 |

**Figures Captions:**

**Fig. 1** XRD pattern of powder at room temperature.

**Fig. 2** Observed (circle), Calculated (continuous lines), and difference (bottom of the figure) profiles in the 2θ range 20-80 degree for the given composition.

**Fig. 3** Scanning electron micrographs of sintered composition

**Fig. 4** Frequency dependence of the (a) ε' and (b) ε'' of the material at various temperatures.

**Fig. 5** Temperature dependence of the (a) ε' and (b) ε'' of the material at various frequencies.

**Fig. 6** The Arrhenius plots corresponding to $\varepsilon''$ and $M''$.

**Fig. 7** Frequency dependence of the $M''$ of the material at various Temperatures.

**Fig. 8** Scaling behavior of M″ at various temperatures for the present composition

**Fig. 9** Frequency spectra of the conductivity for given composition at various temperatures.

**Fig. 10** Frequency dependence of normalized peaks of $M''$ and tanδ for the material at 333K.

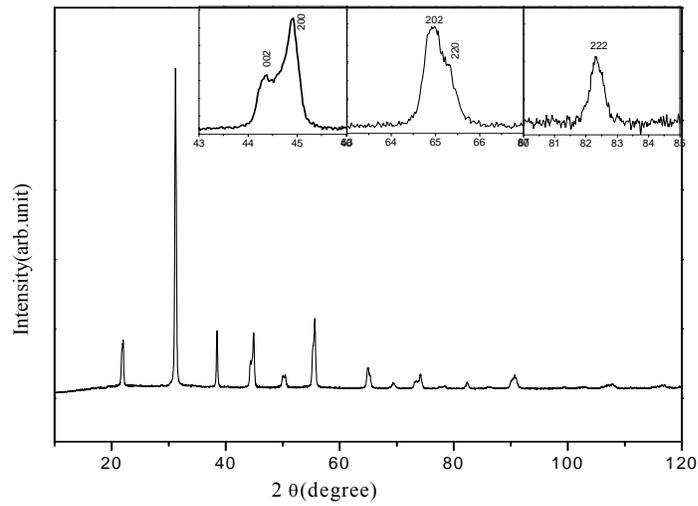

**Fig. 1** XRD pattern of powder at room temperature.

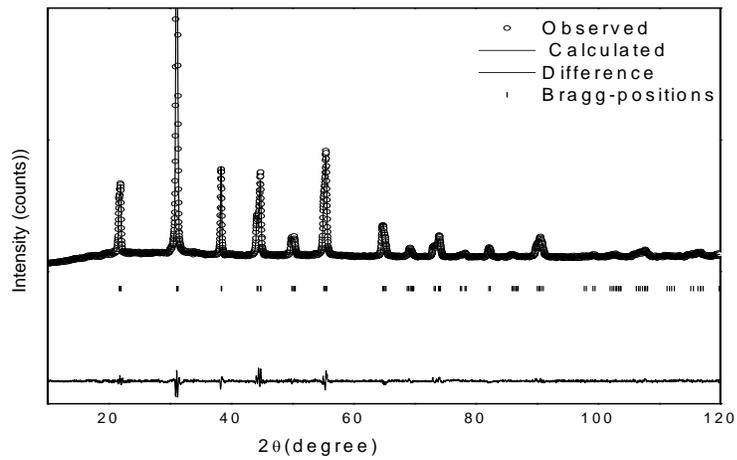

**Fig. 2** Observed (circle), Calculated (continuous lines), and difference (bottom of the figure) profiles in the 2θ range 20-80 degree for the given composition.

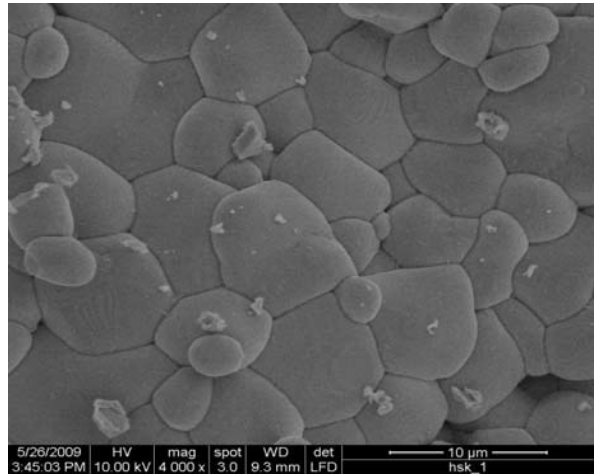

**Fig. 3** Scanning electron micrographs of sintered composition

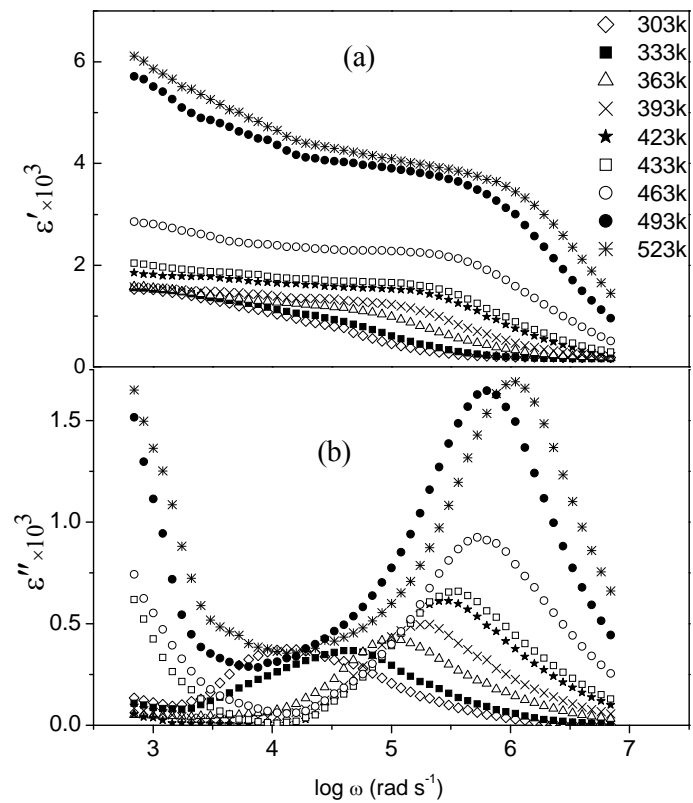

**Fig. 4** Frequency dependence of the (a) ε' and (b) ε'' of the material at various temperatures.

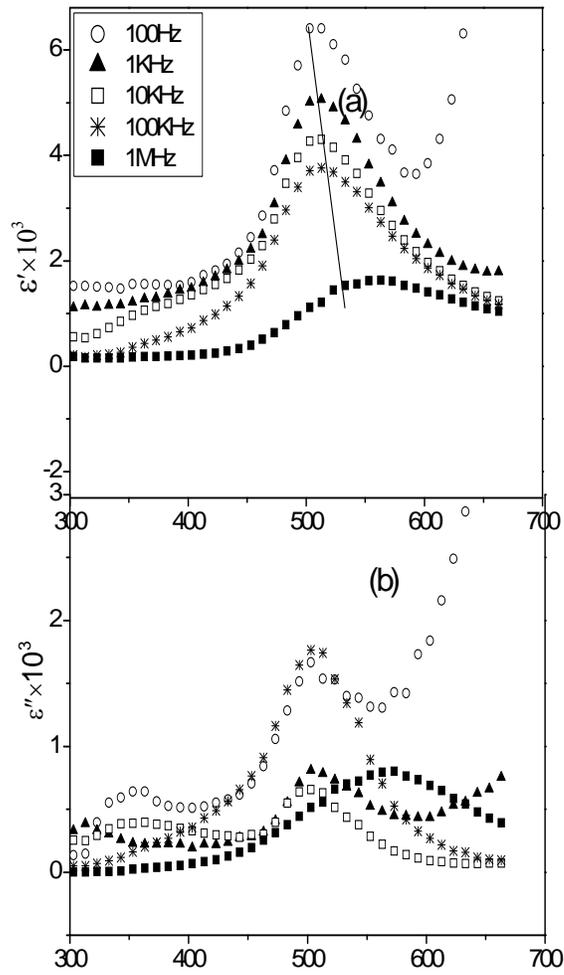

**Fig. 5** Temperature dependence of the (a) ε' and (b) ε'' of the material at various frequencies.

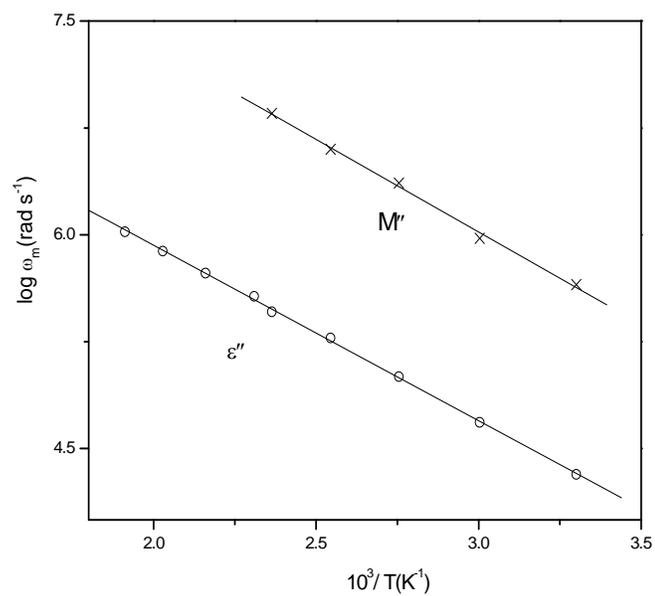

**Fig. 6** The Arrhenius plots corresponding to $\varepsilon''$ and $M''$.

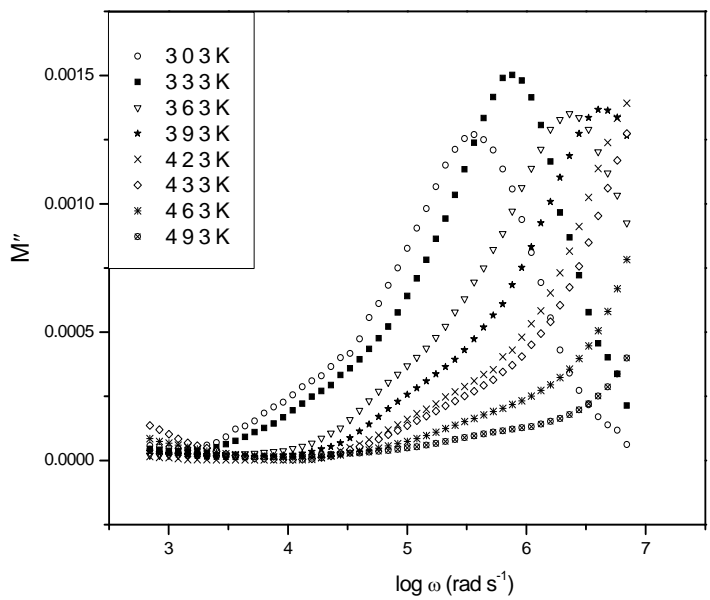

**Fig. 7** Frequency dependence of the $M''$ of the material at various Temperatures.

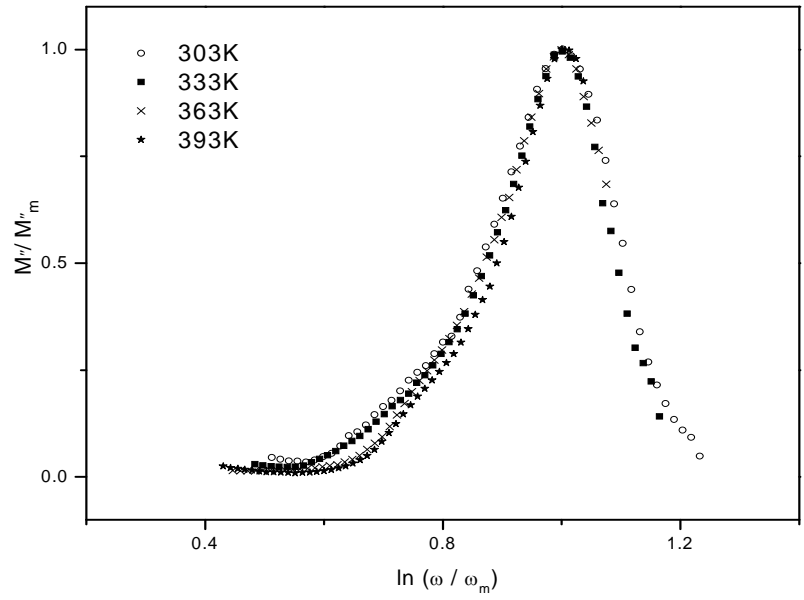

**Fig. 8** Scaling behavior of M″ at various temperatures for the present composition

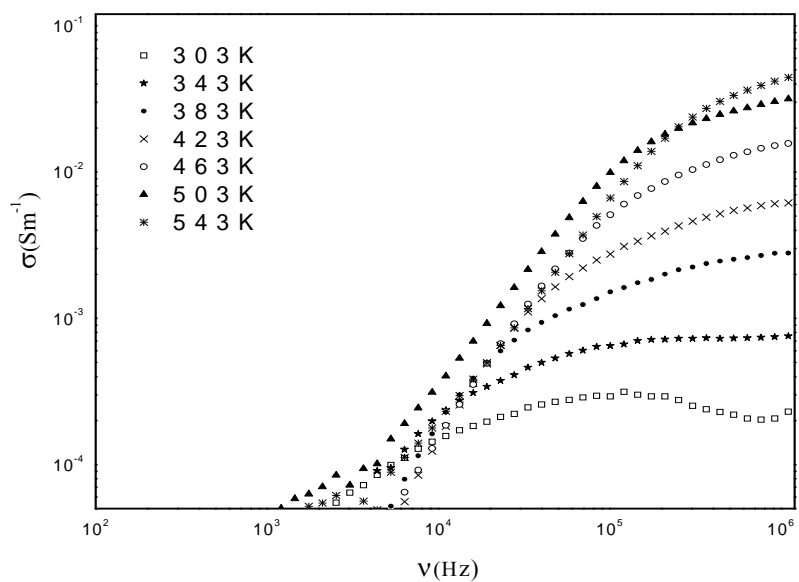

**Fig. 9** Frequency spectra of the conductivity for given composition at various temperatures.

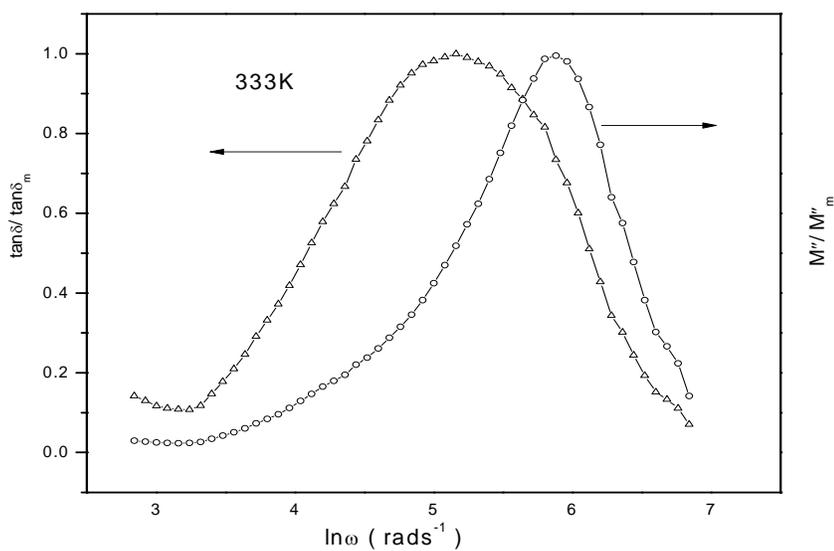

**Fig. 10** Frequency dependence of normalized peaks of M″ and tanδ for the material at 333K.